\documentclass[12pt,english,floatfix,showkeys,superscriptaddress,aps,prd,preprint]{revtex4}
\usepackage[utf8]{inputenc}
\usepackage[english]{babel}
\usepackage[T1]{fontenc}
\usepackage{lmodern}
\usepackage{amsmath}
\usepackage{amssymb}
\usepackage{graphicx}
\usepackage{float}
\usepackage{esint}
\usepackage{longtable}
\usepackage{dcolumn}
\usepackage{babel}
\usepackage{csquotes}
\usepackage{color}
\usepackage{slashed}
\usepackage{simplewick}
\usepackage[utf8]{inputenc}                   
\usepackage{amsmath,latexsym}

\usepackage{hyperref}
\hypersetup{
    colorlinks,
    citecolor=blue,
    filecolor=green,
    linkcolor=purple,
    urlcolor=red,
}

\usepackage{slashed}

\usepackage{hyperref}
\hypersetup{colorlinks,breaklinks,
			citecolor=[rgb]{0,0.0,1.0},
            urlcolor=[rgb]{0.0,0.0,1.0},
            linkcolor=[rgb]{0,0.5,0.9}}

\begin{document}

\title{Probing the Solar System for Dark Matter Using the Sagnac Effect}

\author{A. D. S. Souza}
\email{antonia.daniele@aluno.uece.br}


\author{C. R. Muniz}
\email{celio.muniz@uece.br}


\author{R. M. P. Neves}
\email{raissa.pimentel@uece.br}
\affiliation{Universidade Estadual do Cear\'a (UECE), Faculdade de Educa\c{c}\~ao, Ci\^encias e Letras de Iguatu, Av. D\'ario Rabelo s/n, Iguatu - CE, 63.500-00 - Brazil.}


\author{M. B. Cruz}
\email{messiasdebritocruz@servidor.uepb.edu.br}
\affiliation{Universidade Estadual da Para\'iba (UEPB), \\ Centro de Ci\^encias Exatas e Sociais Aplicadas (CCEA), \\ R. Alfredo Lustosa Cabral, s/n, Salgadinho, Patos - PB, 58706-550 - Brazil.}


\date{\today}

\begin{abstract}

This study investigates the potential of the Sagnac Effect for detecting dark matter in the Solar System, particularly within the Sun. Originating from the relative delay and interference of light beams traveling in opposite directions on rotating platforms, the effect can account for how varying gravitational conditions affect its manifestation. We analyze the Sagnac time in two static, spherically symmetric spacetimes: Schwarzschild and one incorporating dark matter, in the form of a perfect fluid. Comparing the relative deviations in Sagnac time calculated for these metrics in the reference frame of satellites orbiting our star, which serve as a rotating circular platform and emit laser beams in opposite directions, with the precision of onboard atomic clocks (about $10^{-11}$), allows us to evaluate the potential for detecting dark matter's gravitational influence through this effect.
\end{abstract}

\keywords{Sagnnac Effect, Solar System, Dark Matter, Schwarzschild Metric.}

\maketitle


\section{Introduction}

In 1913, French physicist Georges Sagnac conducted a groundbreaking experiment that verified the existence of light interference on a rotating platform. His setup involved splitting a beam of light into two paths using a semi-silvered mirror \cite{Sagnac}. These beams traveled along the platform's perimeter in opposite directions, one favoring and the other opposing the platform's rotation. Upon returning to the source, the beams exhibited a time difference due to their differing paths, resulting in a measurable phase shift in the light evidenced by interference fringe patterns.

Sagnac's experiment aimed to explain various optical phenomena within a specific theoretical framework, including the Fresnel-Fizeau experiment that demonstrated the drag of light in a moving medium. Interestingly, around 1910, the prevailing paradigm held onto the concept of absolute space and its associated hypothetical medium for light propagation, the ether. According to Sagnac, the effect he had discovered corroborated this concept and refuted Einstein's Relativity \cite{ruggiero2015sagnac}. Ironically, the Sagnac device, initially conceived to disprove special relativity, is now a key tool in testing general relativity and even its extensions \cite{tartaglia1998,Nandi:2000xt,Nandi2002,ruggiero2005,Sultana:2013apa,turimov2023propagation,karimov2018,benedetto2019,iovane2020,feleppa2021,Ziaie:2022zmz}. Since its discovery, the Sagnac effect has been extensively studied and found to have numerous crucial applications in various fields. Examples include fiber optic gyroscopes used in navigation systems for airplanes, ships, and missiles, and ring laser gyroscopes employed to measure Earth's rotation. However, it is important to note that experiments by Wang et al. \cite{Aipin} have raised questions about the pivotal role of rotation in the Sagnac effect, suggesting that the origin of the Sagnac delay appears to be different, as indicated in reference \cite{Bhadra}.

The Sagnac effect results in a proper time difference between two observers moving in opposite directions along closed paths, and has been validated through atomic clocks flown around the Earth. Inspired by this, the authors of Ref. \cite{Hush} introduce a novel interferometer that utilizes a single atomic clock, which would significantly simplify the task of measuring the effect, especially over large distances. In this configuration, the Sagnac effect manifests as a phase shift between trapped atoms in different internal states after they are transported along closed paths without free propagation. This approach leverages the precision of atomic transitions to detect phase shifts induced by the Sagnac effect, offering valuable insights into the influences of rotation, gravity, and, particularly, dark matter on quantum systems, which is of interest in our study.

The elusive nature of dark matter persists as one of the most intriguing unsolved mysteries in physics. Nevertheless, this invisible substance is believed to constitute approximately five-sixths of all matter in the universe \cite{aghanim2020planck}. Its undeniable presence has been substantiated by numerous astrophysical and cosmological observations over the past decades \cite{zwicky1933rotverschiebung,bertone2018history}. For instance, galaxy rotation curves, where the observed velocities of stars at their peripheries defy the expected distribution of visible matter, point towards the existence of invisible dark matter \cite{persic1996}. While estimates suggest its mass ranges from large-scale astrophysical structures to a minimum limit influenced by quantum pressure (around $10^{-22}$ eV), its composition and particle nature remains elusive \cite{randall2018}. Identifying and experimentally testing potential dark matter candidates is a crucial objective in modern physics.

One intriguing hypothesis involves primordial black holes, formed shortly after the Big Bang and potentially still drifting through the cosmos. These differ significantly from stellar and supermassive black holes found at galactic centers. Their gravitational influence on surrounding matter could hint at their presence \cite{pablo2021}. Stephen Hawking famously proposed that newly forming stars could, under rare circumstances, capture these primordial black holes \cite{hawking1971}, offering insights into dark matter, which also could be produced by the evaporation of these objects during its lifetime \cite{cheek2022}. Stars harboring such black holes at their cores, known as Hawking stars, could exhibit surprisingly extended lifespans. Even our Sun could potentially harbor a black hole as massive as Mercury without our knowledge \cite{bellinger2023}. This stability arises from the outward energy flow from nuclear fusion counteracting the gravitational collapse tendency. Identifying a Hawking star would have profound implications, potentially unlocking the dark matter mystery and confirming the existence of primordial black holes.

Another possibility lies in extensions of the Standard Model, proposing new types of particles. Weakly interacting massive particles (WIMPs) are a prominent candidate, envisioned as neutral, non-relativistic particles with masses ranging from a few GeV to $10^3$ GeV and weak interactions with Standard Model particles \cite{garret2010}. These hypothetical particles could be gravitationally pulled into the Sun, accumulating and annihilating into Standard Model particles like neutrinos. Detecting such phenomena, while not achieved yet by the IceCube Collaboration, could set constraints on dark matter annihilation into neutrinos \cite{alejandro2021}. Moreover, recent research suggests that dark photons may play a significant role in understanding dark matter, particularly in the context of the Sun \cite{Emken, Feng}

Motivated by the ubiquity of dark matter within and beyond our galaxy, alongside recent advancements in detection methods, our investigation focuses on its feasibility within the framework of General Relativity. Our specific objective is to analyze the gravitational influence of dark matter hidden within the Sun's core on outer measurements of the Sagnac effect. This involves employing precision clocks installed on satellites orbiting our star for observation and analysis related to the phenomenon. 

Initially, we apply this framework to the Schwarzschild exterior metric to illustrate the method and later extend it to study a specific dark matter model. Thus, we calculate the relative shift in the arrival time intervals of light beams emitted in opposite directions along an orbiting satellite network that travels in the same direction. We will conduct this analysis in both the presence and absence of dark matter and compare the results with the current precision of onboard clocks. The selected metric describes the spacetime around a source surrounded by a halo of dark matter, modeled as a perfect fluid (PFDM) with a mass equivalent to up to 0.1\% of the Sun's mass. 

Despite its origins in galactic studies \cite{li2012galactic}, the use of perfect fluid dark matter (PFDM) is a simplified approximation of a more complex reality, chosen as a preliminary model for this study. In this work, we assume that the Sun is surrounded by an envelope of approximately collisionless dark matter particles, with a density profile $\rho\propto r^{-3}$, consistent with the outer tail of the Navarro-Frenk-White (NFW) profile \cite{Charles}. This approach allows PFDM to maintain theoretical consistency when applied to astrophysical sources like stellar-sized black holes and stars, as supported by several recent studies \cite{sanjar2021,ma2021,kimet2021,xiao2023,rizwan2023,quiao2023,ghosh2023,Sharif,Sadeghi,Anjum,Deng,Fard,Atamurotov,Abdusattar}. Therefore, our assumption supports the validity of PFDM for describing objects similar to the Sun in both mass and size. Moreover, the proposed experimental setup allows one to validate the model's applicability at this scale.

The work is structured as follows: In Section \ref{sec_2}, we do a brief overview of the Sagnac effect. In Section \ref{sec_3}, we calculate the Sagnac time for the Schwarzschild spacetime and the dark matter halo scenario, followed by an analysis of the relative variations in these values compared between themselves. These results are then used to assess the viability of measuring these deviations using atomic clocks, considering the limitations of current technology and experimental sensitivity. Finally, Section \ref{sec_4} presents our conclusions and final considerations.
\section{The Sagnac Effect} \label{sec_2}

\begin{figure}[h]
    \centering
    \includegraphics[scale=0.3]{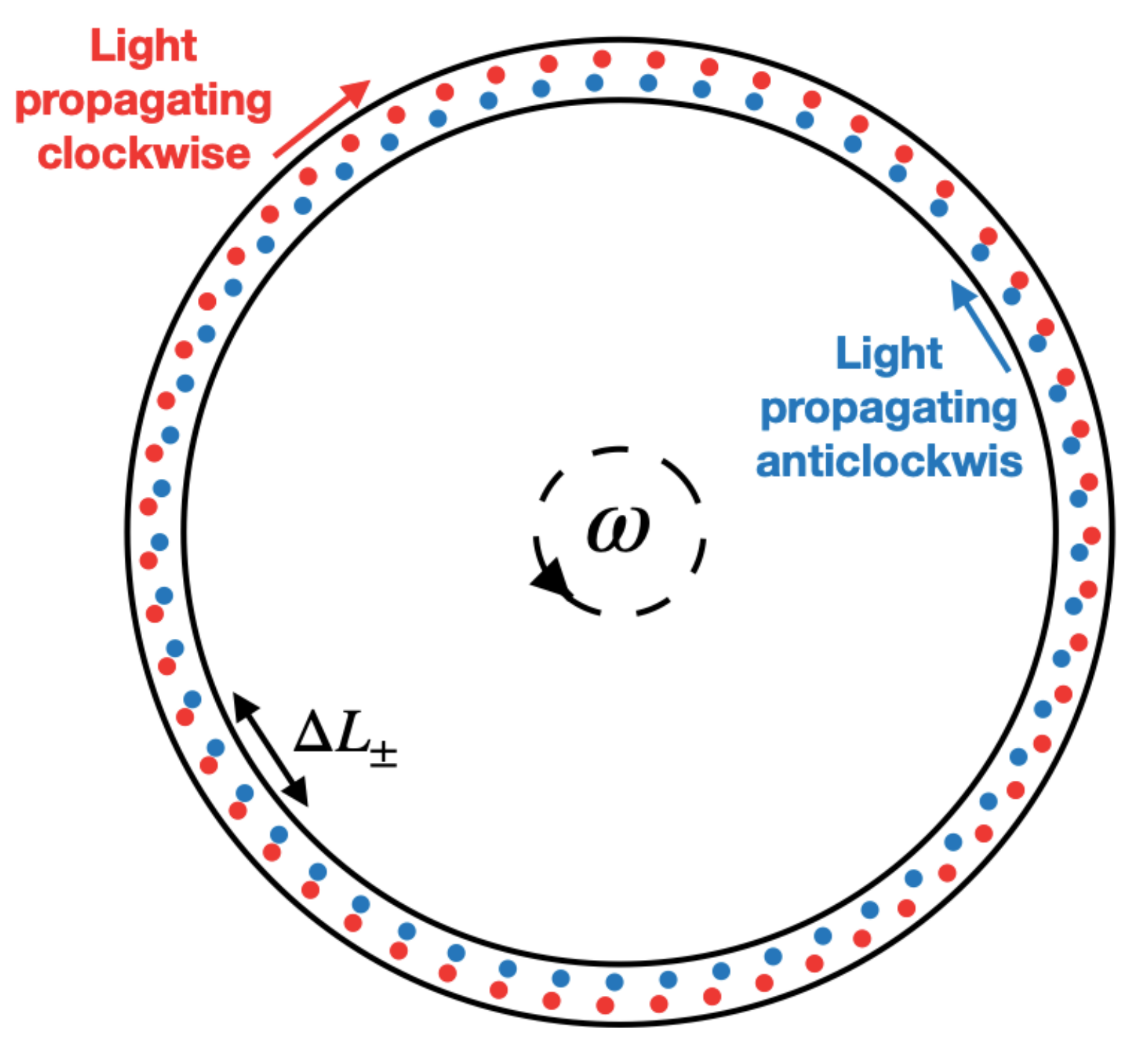}
    \caption{Representation of the Sagnac interferometer, with light beams traveling in opposite directions along the perimeter of a circular platform that rotates with angular velocity $\omega$.}
    \label{sagnac_effect}
\end{figure}

As mentioned earlier, the Sagnac effect involves a source emitting a light beam that splits into two beams traveling in opposite directions along a rotating circular platform. These beams eventually return to the source at different times, as represented in Figure \ref{sagnac_effect}. Considering two light beams traveling along the perimeter of the platform, which has a length of $L$. The beam rotating in the same direction as the platform's rotation will reach the emitter at a distance $\Delta L_{+}$ and a time $t_+$, given by:
\begin{equation} \label{dis_tem_+}
    \Delta L_+ = v t_+ \ \ \ \ \ \ \text{and} \ \ \ \ \ t_+ = \frac{L + \Delta L_{+}}{c}.
\end{equation}
Here, $v$ represents the linear velocity of the source. Consequently, the total distance traveled by the light (assuming its velocity is constant $c$) is the sum of the platform's length $L$ and the additional distance $\Delta L_{+}$.

Similarly, the light beam traveling against the rotation of the platform has a distance $\Delta L_{-}$ associated with a time $t_-$, given by:
\begin{equation} \label{dis_tem_-}
    \Delta L_{-} = v t_- \ \ \ \ \ \ \text{and} \ \ \ \ \ t_- = \frac{L - \Delta L_{-}}{c}.
\end{equation}
By utilizing Eqs. \eqref{dis_tem_+} and \eqref{dis_tem_-}, we can determine the time interval $\Delta t$ between the arrival of the two beams at the source. This interval depends on the platform's length $L$, the speed of light $c$, and the platform's speed $v$. The resulting expression is 
\begin{eqnarray} \label{time_interval}
    \Delta t =  t_+- t_- = \frac{2Lv}{c^2-v^2} .
\end{eqnarray}

In order to simplify Eq. \eqref{time_interval}, we assume the light travels along a circular path with radius $R$. This allows us to express the platform's length as $L = 2\pi R$. Furthermore, we can relate the linear velocity $v$ to the angular velocity $\omega$ through the equation $v= \omega R$. Therefore, the Sagnac time in terms of the constant angular velocity is given by
\begin{equation} \label{time_interval_approx}
    \Delta t = \frac{4 \pi R^2 \omega}{c^2 - v^2} \approx 4 \frac{A \omega}{c^2}.
\end{equation}
Here, $A = \pi R^2$ represents the area of the circular platform. Importantly, we can assume $c \gg v$ in most practical situations.

Drawing on the insights gained from analyzing the Sagnac effect within the framework of Special Relativity \cite{tartaglia2015sagnac}, we can further extend the concept of the Sagnac experiment by considering the Minkowski spacetime line element in spherical coordinates:
\begin{eqnarray} \label{minkowski_metric}
    ds^2 = c^2 dt^2 - dr^2 - r^2 d\theta^2 - r^2 \sin^2\theta d\phi^2.
\end{eqnarray}
At this point, we assume the rotation occurs in the equatorial plane, meaning $\theta = \pi/2$ and the radius is constant, $R$. We then perform a metric transformation to a rotating frame with angular velocity $\omega$ and angular displacement $\phi = \phi_0 - \omega t$. Considering the differential form $d\phi = d\phi_0 - \omega dt$, the metric of Eq. \eqref{minkowski_metric} transforms into:
\begin{eqnarray} \label{minkowski_metric_equa_plane}
     ds^2 = (c^2 - R^2 \omega^2)dt^2 - R^2 d\phi^2 + 2R^2 \omega dt d\phi.
\end{eqnarray}
In this form, the subscript $0$ in the coordinate $\phi$ has been suppressed. For further analysis, equation \eqref{minkowski_metric_equa_plane} can be rewritten as $ds^2 = g_{t} dt^2 + g_{\phi \phi} d\phi^2 + 2g_{t \phi} d\phi dt$, with metric components given by: $g_{tt} = c^2 -  R^2 \omega^2$, $g_{\phi \phi} = - R^2$ and $g_{t \phi} = R^2 \omega$.

The trajectory of light in spacetime is defined by setting $ds=0$. To find the time difference between the arrival of the two light beams at the source, we solve the equation for $dt$ and obtain the following result:
\begin{eqnarray} \label{inf_time_relat}
    dt = \frac{- g_{t \phi} \pm \sqrt{{g_{t \phi}} ^2 - g_{tt} g_{\phi \phi}}}{g_{tt}} d\phi.
\end{eqnarray}
Considering the positive root and integrating along the direction with $d\phi > 0$, we obtain the circulation time of light traveling in the same direction as the platform rotation as:
\begin{eqnarray} \label{time_relat_+}
    t_+ = - \int_{0}^{2 \pi} \frac{g_{t \phi}}{g_{tt}} d\phi + \int_{0}^{2 \pi} \frac{\sqrt{{g_{t \phi}}^2 - g_{tt} g_{\phi \phi}}}{g_{tt}} d\phi.
\end{eqnarray}
For the opposite direction, where $d\phi < 0$, the corresponding time is given by:
\begin{eqnarray} \label{time_relat_-}
    t_- = \int_{0}^{2 \pi} \frac{g_{t \phi}}{g_{tt}} d\phi + \int_{0}^{2 \pi} \frac{\sqrt{{g_{t \phi}}^2 - g_{tt} g_{\phi \phi}}}{g_{tt}} d\phi.
\end{eqnarray}
Consequently, from Eqs. \eqref{time_relat_+} and \eqref{time_relat_-}, the time interval $\Delta t$ is
\begin{eqnarray} \label{time_interval_relat}
    \Delta t = t_+ - t_- = - 2 \int_{0}^{2 \pi} \frac{g_{t \phi}}{g_{tt}} d\phi.
\end{eqnarray}

However, expressing the Sagnac time for an observer positioned on the disk is crucial. This observer measures the invariant proper time $\Delta \tau$, which is given by:
\begin{eqnarray} \label{proper_time}
    \Delta \tau = \sqrt{g_{tt}} \Delta t.
\end{eqnarray}
Substituting the expression from Eq. \eqref{time_interval_relat} into Eq. \eqref{proper_time}, we obtain
\begin{eqnarray} \label{Sagnac_time}
    \Delta \tau = - 2 \sqrt{g_{tt}} \int_{0}^{2 \pi} \frac{g_{t \phi}}{g_{tt}} d\phi = -4 \pi \frac{g_{t \phi}}{\sqrt{g_{tt}}}. 
\end{eqnarray}
Observe that the static metric (coefficients constant in time) simplifies integration concerning $d\phi$.

The Eq. \eqref{Sagnac_time} provides the proper Sagnac time as a function of the metric coefficients. In the next section, we will illustrate its use by calculating $\Delta \tau$ for two spherically symmetric spacetimes: the Schwarzschild metric (representing a spherical distribution of baryonic matter) and the metric with dark matter content. We will analyze the possibility of detecting the influence of both baryonic matter and dark matter by comparing the Sagnac time deviation with the accuracy of current atomic clocks.

\section{Applying the Sagnac Effect to Probing Baryonic Mass and Dark Matter in the Solar System} \label{sec_3}

In this section, we will explore the potential of the Sagnac effect to detect ordinary baryonic mass and dark matter within the Sun. Thus, we analyze the Sagnac time for two different spacetimes: Schwarzschild (without dark matter) and Schwarzschild which incorporates dark matter. Also, we will compare the relative Sagnac time deviations in each spacetime with the accuracy of current atomic clocks of GPS, typically $10^{-11}$ \cite{bernardo2021}. This comparison will serve as a crucial threshold for discerning potential detection.

\subsection{Schwarzschild Spacetime}

The Schwarzschild metric, a solution to Einstein's field equations in vacuum, describes the spacetime around a spherically symmetric and static mass distribution, like a star or black hole. We will calculate the Sagnac time measured by an observer within this spacetime. The solution in spherical coordinates is given by \cite{carroll2019spacetime}:
\begin{eqnarray} \label{schw_metric}
    ds^2 = \left(1 - \frac{2GM}{c^2 r}\right) c^2 dt^2 - \left(1 - \frac{2GM}{c^2 r}\right)^{-1} dr^2 - r^2 \left(d\theta^2 + \sin^2 \theta d\phi^2 \right),
\end{eqnarray}
where the mass of the gravitational source, $M$, stretches time and warps radial distances due to its gravity.

Next, we will focus on rotational motion in the equatorial plane $\theta = \pi/2$ at a constant radius $R$. Then, we perform a transformation in the metric to a rotating reference frame with constant angular velocity $\omega$. The angular displacement in this frame is given by $\phi = \phi_{0} - \omega t$. This transformation leads to the following form of the metric from Eq. \eqref{schw_metric}:
\begin{eqnarray} \label{transf_schw_metric}
    ds^2= \left(1 - \frac{2GM}{c^2 R} - \frac{R^2 \omega^2}{c^2}\right) c^2 dt^2 - R^2 d\phi^2 + 2 \frac{R^2 \omega}{c} cdt d\phi,
\end{eqnarray}
where, once again, the subscript 0 in the coordinate $\phi$ has been omitted. Notice that the spacetime is modified by centrifugal effects, represented by the terms depending on $\omega$.

The metric components resulting from Eq. \eqref{transf_schw_metric} are as follows: 
\begin{eqnarray} \label{compont_schw_metric}
    g_{tt}= 1 - \frac{2GM}{c^2 R} - \frac{R^2 \omega^2}{c^2}, \ \ \ 
    g_{\phi \phi} = - R^2 \ \ \ \ \text{and} \ \ \ \ g_{t \phi} = \frac{R^2 \omega}{c}.
\end{eqnarray}

Therefore, substituting the metric components of Eq. \eqref{compont_schw_metric} into Eq. \eqref{Sagnac_time} for the Sagnac time, we obtain
\begin{eqnarray} \label{sagnac_time_schw_metric}
    \Delta \tau_{\text{Schwarzschild}} = - \frac{4 \pi}{c^2} \frac{R^2 \omega}{\sqrt{1 - \frac{2GM}{c^2 R} - \frac{R^2 \omega^2}{c^2}}}.
\end{eqnarray}
As the mass $M$ approaches zero, Eq. \eqref{sagnac_time_schw_metric} converges to the Sagnac time in the flat spacetime described by Eq. \eqref{minkowski_metric_equa_plane}. Therefore, the relative deviation of the Sagnac time for the Schwarzschild metric is given by:
\begin{eqnarray} \label{relat_sagnac_time_schw_metric}
    \Delta = \frac{|\Delta \tau_{\text{Schwarzschild}} - \Delta \tau_{\text{Minkowski}}|}{\Delta{\tau}_{\text{Minkowski}}} = \frac{\left|\frac{1}{\sqrt{1 - \frac{R^2 \omega^2}{c^2}}} - \frac{1}{\sqrt{1 - \frac{2GM}{c^2 R} - \frac{R^2 \omega^2}{c^2}}} \right|}{\frac{1}{\sqrt{1 - \frac{R^2 \omega^2}{c^2}}}}.
\end{eqnarray}
To analyze the Sagnac time for observers on the orbits traversed by the planets Mercury, Venus, and Earth, we consider the following constants: Sun mass, $M_{\odot} = 1.98 \times 10^{30} kg$; light speed, $c = 3.00 \times 10^8$ m/s; and gravitational constant, $G = 6.67 \times 10^{-11}$ m$^3$ kg$^{-1}$ s$^{-2}$. The relative deviations of the Sagnac time for these orbits, whose data were extracted from \cite{NASA}, are calculated using Eq. \eqref{relat_sagnac_time_schw_metric}, and are listed in Table I.
\begin{table}[h!]
\centering
\caption{Relative Sagnac time deviations for the Schwarzschild spacetime.}
\vspace{0.5cm}
\begin{tabular}{lrrr}
\hline
Planet & Orbit radius (R) & Angular velocity ($\omega$) & Relat. dev. ($\Delta$) \\
\hline
Mercury & $5.79 \times 10^{10} m$ & $8.26 \times 10^{-7} rad/s$ & $2.54 \times 10^{-8}$ \\
Venus & $1.08 \times 10^{11} m$ & $3.24 \times 10^{-7} rad/s$ & $1.36 \times 10^{-8}$ \\
Earth & $1.50 \times 10^{11} m$ & $1.99 \times 10^{-7} rad/s$ & $9.75 \times 10^{-9}$  \\
\hline
\end{tabular}
\end{table}

Analyzing Table I, we see that the relative Sagnac time deviations for Mercury, Venus, and Earth are within the limit of current clock accuracy ($10^{-11}$). This indicates that detecting the Sun's mass influence on their spacetime through the Sagnac effect is highly feasible. Thus, such an experiment, besides the classical tests \cite{Sanjeev2022}, remains another confident method for measuring the effects of general relativity in the solar system.

\subsection{Schwarzschild Spacetime with Dark Matter}

Our primary objective is to investigate the presence of dark matter within the Sun and explore its potential detection using the Sagnac effect. To achieve this, we analyze the Sagnac time for a metric that incorporates a parameter specifically related to dark matter, denoted by $\lambda$. We will then calculate the relative deviation for this scenario and compare it to the current accuracy of atomic clocks onboard satellites.

We will adopt the solution initially studied in \cite{li2012galactic}, in which the space-time metric arises from a specific model of dark matter in the presence of an additional scalar field (called phantom field). After this, the model was investigated in several other contexts \cite{sanjar2021,ma2021,kimet2021,xiao2023,rizwan2023,quiao2023,ghosh2023,Sharif,Sadeghi,Anjum,Deng,Fard,Atamurotov,Abdusattar}, denoted as perfect fluid dark matter (PFDM). 

The metric describing the static, spherically symmetric exterior spacetime of such a source surrounded by a perfect fluid of dark matter is given by:
\begin{eqnarray} \label{schw_metric_darkmatter}
    ds^2 = F(r)c^2 dt^2 - F(r)^{-1} dr^2 - r^2 \left(d\theta^2 + \sin^2\theta  d\phi^2 \right),
\end{eqnarray}
where the function $F(r)$ is defined by
\begin{eqnarray}
    F(r) = 1 - \frac{2GM}{c^2 r} + \frac{\lambda}{r}\log{\frac{r}{|\lambda|}}.
\end{eqnarray}
In this metric, $M$ represents the mass of the spherically symmetric baryonic source, and $\lambda$ is the distance scale governing the influence of dark matter. The spacetime metric can be derived from Einstein's equations considering the particular case of a perfect fluid with an isotropic density profile generically given by \cite{rossi2024}
\begin{equation}\label{DMProfiles}
    \rho(r)=\frac{\rho_0}{\sum_{n=0}^3 a_n (r/r_0)^n},
\end{equation}
which corresponds to various models of galactic dark matter distribution, depending on whether it is inferred from numerical simulations of collision-less particle clustering or observed rotation curve analysis. In our analysis, we will consider the simplest case in which $a_0=a_1=a_2=0$. Hence, we can define a typical mass of dark matter $k=\rho_0 r_0^3/a_3$, which determines the parameter $\lambda$ via
\begin{eqnarray} \label{dark_matter_param}
      \lambda = \dfrac{8 \pi G k}{c^2}.
\end{eqnarray}
Here, the parameter $k$ represents a local dark matter mass associated with the Sun. In the limit where $\lambda$ approaches zero ($\lambda \to 0$), the effects of dark matter vanish. Consequently, the spacetime metric reduces to the Schwarzschild metric, recovering all the results obtained in the previous subsection.

Now, we consider the same conditions as before: rotational motion in the equatorial plane with a constant radius, i.e., with $\theta = \pi/2$ and $r=R$. We transform the metric to a rotating reference frame with constant angular velocity $\omega$ and angular displacement $\phi = \phi_0 - \omega t$. This leads to the following metric:
\begin{eqnarray} \label{dark_matter_metric_perf}
    ds^2 = \left(1 - \frac{2GM}{c^2 R} + \frac{\lambda}{R}\log {\frac{R}{|\lambda|}} - \frac{R^2 \omega^2}{c^2}\right) c^2 dt^2 - R^2 d\phi^2 + 2 \frac{R^2 \omega}{c} c dt d\phi.
\end{eqnarray}
In Eq. \eqref{dark_matter_metric_perf}, we observe that the metric components $g_{\phi \phi}$ and $g_{t \phi}$ are identical to those of the Schwarzschild metric in Eq. \eqref{transf_schw_metric}. However, the $g_{tt}$ component is modified by the presence of dark matter and is given by
\begin{eqnarray} \label{g_dark_matter}
    g_{tt} = 1 - \frac{2GM}{c^2 R} + \frac{\lambda}{R}\log {\frac{R}{|\lambda|}} - \frac{R^2 \omega^2}{c^2}.
\end{eqnarray}

Therefore, by using Eq. \eqref{Sagnac_time} for the Sagnac time and substituting Eq. \eqref{g_dark_matter} for the relevant metric components, specifically $g_{t \phi}$, from Eq. \eqref{compont_schw_metric}, we obtain the following expression for the Sagnac time:
\begin{eqnarray} \label{prop_time_dark_matter}
    \Delta \tau_{\text{Dark matter}} = - \frac{4 \pi}{c^2} \frac{R^2 \omega}{\sqrt{1 - \frac{2GM}{c^2 R} + \frac{\lambda}{R}\log{\frac{R}{|\lambda|}}   - \frac{R^2 \omega^2}{c^2}}}.
\end{eqnarray}
Consequently, the relative deviation of the Sagnac time due to the dark matter term can be calculated using Eq. \eqref{prop_time_dark_matter}. This gives us:
\begin{eqnarray} \label{dev_Sagnac_time_dark_matter}
    \Delta = \frac{|\Delta \tau_{\text{Dark matter}} - \Delta \tau_{Schwarzschild}|}{\Delta \tau_{Schwarzschild}} = \left|1 - \frac{\sqrt{1- \frac{2GM}{c^2 R} - \frac{R^2 \omega^2}{c^2}}}{\sqrt{1-\frac{2GM}{c^2 R}+\frac{\lambda}{R}\log \frac{R}{|\lambda|} - \frac{R^2 \omega^2}{c^2}}}        \right|.
\end{eqnarray}

To analyze the effects of dark matter on the relative deviations, we assign values to the dark matter mass using the relationship between $\lambda$ and $k$. As before, we assume the experiment is conducted by a group of satellites traveling at distinct points along the orbits of Mercury, Venus, and Earth, in the same direction. We estimate the dark matter content within the Sun to be approximately a few fractional percentages of its total mass. This estimation is derived from the average dark matter density in the vicinity of the Solar System ($\rho_{DM} \approx 0.44 \times 10^{-21} kg/m^3$) \cite{salucci2010} and the estimated diameter of the protoplanetary cloud that contributed to its formation, between $0.01$ and $0.1$ parsec \cite{pudritz2002}. As a result, we infer a gravitationally collapsed dark matter mass in the Sun's core of approximately $10^{-4}M_{\odot}< k< 10^{-3} M_{\odot}$.

For our analysis, we will consider values up to $10^{-3}$ of the Sun's mass ($M_{\odot}$), as shown in Figure 2, since it's possible that the amount of dark matter inside the Sun or its vicinity varied over time. Considering the example of $k = 10^{-4}$ $M_{\odot}$, we get the characteristic reach of the dark matter as
\begin{eqnarray} \label{lambdas}
      \lambda = \frac{8 \pi G k}{c^2} = 3.39 \ \text{m}.
\end{eqnarray}
\begin{figure}[h]
       \centering
    \includegraphics[scale=0.8]{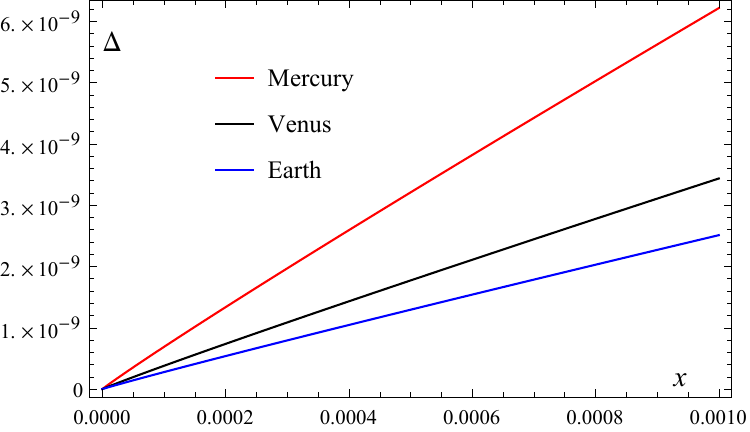}
    \caption{Relative Sagnac time deviation for the orbits of the first three planets in the solar system, plotted as a function of the ratio between dark matter and solar masses, $x=k/M$.}
     \label{fig_ratio_darkmatter_solar_masses}
\end{figure}

\begin{table}[h]
\centering
\caption{Relative Sagnac time deviations for a dark matter mass of $0.01\%$ of the Sun's mass.}
\vspace{0.5cm}
\begin{tabular}{lrrr}
\hline
Planet &Orbit radius ($R$) & Angular velocity ($\omega$)  & Relat. dev. ($\Delta$) \\
\hline
Mercury & $5.79 \times 10^{10} $ m & $8.26 \times 10^{-7} $ rad/s & $6.88 \times 10^{-10}$ \\
Venus & $1.08 \times 10^{11} $ m & $3.24 \times 10^{-7} $ rad/s & $3.79 \times 10^{-10} $ \\
Earth & $1.50 \times 10^{11} $ m & $1.99 \times 10^{-7} $ rad/s & $2.77 \times 10^{-10} $  \\
\hline
\end{tabular}
\end{table}
Based on these results, we calculate the relative deviations using Eq. \eqref{dev_Sagnac_time_dark_matter} and present them in Table II, where once more we used the data collected from \cite{NASA}. Importantly, the relative deviations of the Sagnac time are one order of magnitude larger than the precision of atomic clocks on board satellites. This suggests that the current methods could potentially be used to detect and study the effects of dark matter.

It is worth mentioning that, on considering the upper bound for the dark matter mass within the Solar System, estimated at $\sim 10^{-10} M_{\odot}$ based solely on observations of planetary and spacecraft motions \cite{pitjeva2013}, the relative deviation in the Sagnac Effect, according to Eq. (\ref{dev_Sagnac_time_dark_matter}), decreases to $10^{-14}$ at the Mercury orbit radius, staying significantly below, therefore, the current accuracy of onboard atomic clocks.

We close this Section by stating that the gravitational influence of the planets eventually harboring dark matter on light beams passing nearby does not significantly affect the Sagnac time delay around our star. Using Eq. (\ref{dev_Sagnac_time_dark_matter}) for $\Delta$ and considering Mercury’s data, for instance, with a group of satellites orbiting this planet at $R\approx 2,500$ km and emitting light beams in opposite directions between them, we find that the corresponding relative deviation of the Sagnac time has magnitude order of ($\Delta \sim 10^{-12}$), in the limit of sensitivity, therefore, of the current embarked atomic clocks. This result assumes the same concentration of dark matter within the planet as that used for the Sun’s core, which is unrealistic. In this case, the deviation is two orders of magnitude smaller than those calculated for the Sun.

\section{Conclusion} \label{sec_4}

This study utilized the Sagnac effect to investigate the spacetime geometry of the Solar System, describing a methodology to detect dark matter within the Sun based on its gravitational influence. We employed a simple model treating the outer spacetime of a spherical source surrounded by a dark matter halo as Schwarzschild-like, with dark matter modeled as a perfect fluid (PFDM). Our method for estimating dark matter using the Sagnac effect is theoretical but offers a complementary perspective to existing approaches. While traditional methods like planetary dynamics and gravitational lensing excel on larger scales, they often fail to isolate dark matter effects locally. In contrast, our approach targets local metric perturbations, which may reveal subtle dark matter influences through delicate experiments involving optical or other quantum interferometry.

Our analysis focused on the orbits corresponding to the first three interior planets, examining relative deviations in Sagnac time measured by clocks in satellites following similar trajectories, which travel in the same direction. These satellites function as the circular rotating platform originally designed for the Sagnac effect, from which laser beams are sent in opposite directions to assess such deviations. Our investigation included comparisons with Schwarzschild's spacetime, for which we previously calculated similar deviations compared to Minkowski's spacetime in order to establish a methodological framework.

We explored varying dark matter masses within the Sun, up to $0.1\%$ of the Sun's mass, as illustrated in Figure \ref{fig_ratio_darkmatter_solar_masses}, with a specific focus on the value of $k\sim 0.01\% M_{\odot}$. This yielded a relative deviation in Sagnac time on the order of $10^{-10}$ for the examined orbits, as is indicated in Table II. Our results, highlighted in Tables I and II, demonstrate that these deviations, for both baryonic and non-baryonic matter, surpass the precision of current atomic clocks ($10^{-11}$). This underscores the potential of the Sagnac effect as a valuable tool for detecting and probing spacetime geometry within our Solar System, focusing on the deviation relative to the Schwarzschild metric, which accounts for baryonic matter. This distinction is crucial as it allows us to isolate the potential influence of dark matter. As a bonus, the proposed methodology also offers the capability to test the PFDM model at the local level.

It is important to mention that the Sun's differential spinning complicates the modeling of its rotational effects; however, we can approximate these using the Kerr metric in the slow-rotation limit to account for the Lense-Thirring effect on the Sagnac time. In this context, the relative deviation \(\Delta\) in Eq. (\ref{dev_Sagnac_time_dark_matter}) should include the term $\frac{4 a G M \omega}{c^4 R}$ under both square roots, considering the linear approximation of the rotational Kerr parameter \(a\) for the Sun \cite{Ziaie:2022zmz}. For the orbit of Mercury, this term is around \(10^{-17}\) and even smaller for other planets, making it negligible for our analysis.

It is also worth emphasizing that our work is based on General Relativity (GR), which remains the best-tested gravitational theory to date. We chose to model dark matter within the energy-momentum tensor, aligning with its role as a potential extension of the standard model of elementary particles. Alternatively, to distinguish whether the Sagnac time delay could result from modified gravity theories rather than dark matter, we could introduce metric terms representing modifications to GR (such as massive gravity \cite{Ghosh2016}) within the square root in the denominator of Eq. (25), replacing the logarithmic term for dark matter. We could then compare these results with those of Eq. (25) to assess the source of the Sagnac effect.

This study enhances our understanding of dark matter and promotes further exploration of the Sagnac effect as a detection method. While we acknowledge the experimental challenges and limitations in current sensitivity, our work aims to inspire future advancements in detection techniques, as the one described in \cite{Hush}, which requires solely an atomic clock. We also recognize that incorporating more complex and realistic density profiles, such as those derived from the Navarro-Frenk-White (NFW) model or other empirical distributions, could enhance our theoretical results. Ongoing research in this field is essential for improving our understanding and enhancing methodologies for dark matter detection in the local regions of our galaxy.

Finally, we would like to highlight a recently published article that closely aligns with our work \cite{Tsai:2021lly}. The authors discuss how recent advancements in quantum sensors, particularly atomic clocks, open new possibilities for dark matter searches, especially for ultralight dark matter bound to the Sun. These space-based atomic clocks, orbiting our star, could directly probe the solar system's interior, providing strong constraints on dark matter halos through the detection of ultralight dark matter interactions with electron, photon, and gluon fields, utilizing both current and upcoming atomic, molecular, and nuclear clocks.

\section*{Acknowledgments}
\hspace{0.5cm} CRM thanks the Conselho Nacional de Desenvolvimento Cient\'{i}fico e Tecnol\'{o}gico (CNPq), Grants no. 308268/2021-6.


\end{document}